\documentclass[]{acoust_arXiv}
\def\header{Intensity estimation with Tight-Frame array} 

\title{Acoustic intensity estimation using cardioid microphone pairs in tight-frame configurations}
\subtitle{}

\author{Akira Omoto \thanks{omoto@design.kyushu-u.ac.jp}}

\inst{Faculty of Design, Kyushu University}

\usepackage{xcolor}

\abst{This paper investigates acoustic intensity estimation using pairs of cardioid microphones based on the cardioid-cardioid (C-C) method. 
Unlike conventional pressure-difference techniques, the C-C method is intrinsically less sensitive to the relationship between microphone spacing and acoustic wavelength. 
However, practical microphones inevitably deviate from ideal cardioid directivity, producing direction-dependent estimation errors.
To improve robustness against such errors, a measurement framework based on spherical tight-frame microphone configurations is proposed. 
Directional intensity components measured along multiple axes are combined to reconstruct the three-dimensional acoustic intensity vector. 
Furthermore, directivity errors are represented using Legendre polynomial and spherical harmonic expansions, and a geometry-dependent leakage metric is introduced to quantify the error-suppression capability of different microphone arrangements.
Theoretical analysis and numerical simulations demonstrate that tight-frame configurations effectively suppress direction-dependent errors through geometric averaging. 
The proposed leakage metric provides a qualitative indication of microphone directivity imperfections on the reconstructed intensity vector. 
The results further indicate that accurate wide-band acoustic-intensity estimation can be achieved even with relatively large microphone spacings, which are generally impractical in conventional pressure-difference approaches.
The proposed framework provides a physically interpretable and practically useful approach for acoustic intensity measurement using directional microphone arrays.
}

\kword{Sound intensity, Cardioid microphone, Tight-frame}

\mladdress{Shiobaru 4-9-1, Minami, Fukuoka 815-8540, Japan} 

\class{8} 

\recommend{2}  

\newcommand{\bs}[1]{\mbox{\boldmath $#1$}}
\newcommand{\ii}{\mathrm{i}}

\begin{document}
\maketitle

\section{Introduction}

Acoustic intensity, defined as the rate of acoustic energy flow through a unit area normal to the direction of propagation, is one of the most important quantities for analyzing sound fields. Since active acoustic intensity is given by the ``real part'' of the product of sound pressure and particle velocity, its measurement requires simultaneous estimation of these two quantities.

Among the various measurement techniques, the pressure-pressure (P-P) method is the most widely established. In this approach, two closely spaced omnidirectional microphones are used to estimate sound pressure from their sum and particle velocity from a finite-difference approximation of the pressure gradient. The theoretical basis, practical limitations, and calibration procedures of the P-P method have been extensively investigated and are well documented in the literature \cite{Fahy1995}. Commercial measurement systems based on this principle are also widely available. Alternative approaches employ particle-velocity sensors, such as hot-wire probes or Microflown sensors, which directly measure acoustic particle velocity \cite{deBree2003,Jacobsen2005}.

In spatial-audio and room-acoustic analysis, active-intensity vectors derived from first-order B-format microphone signals have also been used to estimate time- and frequency-dependent directions of arrival and diffuseness, as demonstrated in spatial impulse response rendering (SIRR) \cite{Merimaa2005}, an approach later extended to Directional Audio Coding (DirAC) \cite{Pulkki2007}.

Microphone-array-based approaches have also been proposed, including methods for calculating acoustic intensity using Kirchhoff-Helmholtz integral formulations combined with numerical quadrature schemes \cite{Hacihabiboglu2014}, spherical-harmonic-domain methods that construct a pseudointensity vector from zero- and first-order eigenbeams for three-dimensional source localization \cite{Jarrett2010}, and higher-order generalizations of acoustic intensity \cite{Herzog2021}.

Another class of methods utilizes pairs of cardioid microphones instead of omnidirectional microphones. In this approach, often referred to as the cardioid-cardioid (C-C) method, two oppositely oriented cardioid microphones are combined to estimate sound pressure and particle velocity from their sum and difference signals. The concept dates back to the work of Bauer \cite{Bauer1968} and has subsequently been investigated and improved by several researchers, particularly Hanyu and Hoshi, from both theoretical and practical viewpoints \cite{Hoshi2018,Hanyu2024,Hanyu2026}. Because directional sensitivity is incorporated into the sensing mechanism, the C-C method is intrinsically less dependent on the relationship between microphone spacing and acoustic wavelength than conventional pressure-difference techniques.

Nevertheless, practical implementations of the C-C method are inevitably affected by imperfections of microphone directivity. Real microphones deviate from ideal cardioid characteristics, and these deviations introduce direction-dependent errors into the estimated intensity. When multiple microphone pairs are combined to reconstruct a three-dimensional intensity vector, the manner in which such directional errors propagate through the reconstruction process becomes an important issue.

In a conventional three-dimensional intensity probe, three orthogonal measurement axes are typically employed to estimate the Cartesian components of the intensity vector. More generally, however, intensity can be measured along an arbitrary set of directions and subsequently combined to reconstruct the underlying three-dimensional vector. This observation naturally leads to an overdetermined measurement framework in which directional measurements are obtained along many axes and averaged in a geometrically meaningful manner.

The present study investigates such a framework using microphone configurations based on tight frames. A tight frame is a set of directions possessing a high degree of geometric symmetry and has been studied extensively in frame theory and spherical design theory \cite{Casazza2006,Barg2015,Waldron2018,Hughes2021}. When directional measurements are acquired along a tight frame, the reconstruction of a three-dimensional vector can be expressed as a simple weighted summation of the directional observations. In addition to computational simplicity, the redundancy of the frame provides an averaging effect that is expected to suppress direction-dependent errors.

The motivation of the present work originates from our previous study on diffuseness evaluation using a tight-frame microphone-array configuration \cite{Omoto2026}. In that study, a configuration consisting of twenty-four microphones arranged on a sphere exhibited surprisingly robust performance at high frequencies, even when practical directional microphones were employed. The physical mechanism underlying this robustness, however, has not yet been clarified.

The objective of the present paper is therefore to investigate the error-suppression capability of tight-frame microphone configurations for acoustic-intensity measurement based on the C-C method. Direction-dependent directivity errors are modeled using Legendre polynomial and spherical harmonic expansions, and a geometry-dependent leakage metric is introduced to quantify how individual angular-error components propagate through the reconstruction process. The theoretical framework is validated through numerical simulations using ideal and measured directivity data of a practical cardioid microphone. The results demonstrate that geometrically redundant tight-frame configurations substantially reduce directional estimation errors and enable stable wide-band acoustic-intensity measurements even with relatively large microphone spacings.

The scope of this study is limited to relative comparisons among the evaluated tight-frame configurations under a common reconstruction framework. 
We do not claim optimality over arbitrary non-tight-frame arrays, whose performance would also depend on the microphone geometry, aperture, channel pairing, and decoder design.


\section{Principle of Intensity Estimation Using a Cardioid Pair}

\subsection{Plane-wave formulation}

For simplicity, consider a plane wave arriving in the direction specified by the unit vector

\begin{equation}
\bs{s}
=
\begin{bmatrix}
s_x \;\;&
s_y \;\;&
s_z
\end{bmatrix}^{T}.
\end{equation}
Here, $\bs{s}$ points from the array center toward the source; the plane-wave propagation direction is therefore $-\bs{s}$.

The axis of a pair of oppositely oriented cardioid microphones is represented by

\begin{equation}
\bs{r}
=
\begin{bmatrix}
r_x \;\;&
r_y \;\;&
r_z
\end{bmatrix}^{T}.
\end{equation}

The directivity of an ideal cardioid microphone is given by

\begin{equation}
D(\bs{r},\bs{s})
=
\frac{1}{2}
\left(
1+\bs{s}^{T}\bs{r}
\right)
=
\frac{1}{2}
\left(
1+\cos\theta
\right),
\label{eq:cardioid}
\end{equation}
where $\theta$ denotes the angle between $\bs{r}$ and $\bs{s}$, and
$\bs{s}^{T}\bs{r}=\cos\theta$.

\begin{figure}[t]
\centering
\includegraphics[bb={0 0 73.7892 83.3208},width=0.35\linewidth,pagebox=artbox]{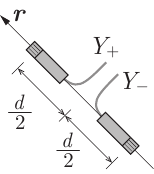}
\caption{Typical microphone arrangement for the C-C method. Although a back-to-back configuration is illustrated, the same principle can also be implemented using two opposed cardioid microphones.}
\label{fig:mic_arrange}
\end{figure}

Consider two cardioid microphones separated by a finite distance $d$ and arranged in opposite directions as shown in Fig. \ref{fig:mic_arrange}.

The microphone positions are defined as
\begin{equation}
\bs{x}_{+}
=
\frac{d}{2}\bs{r},
\qquad
\bs{x}_{-}
=
-\frac{d}{2}\bs{r},
\label{eq:mic_position}
\end{equation}
where the midpoint of the microphone pair is taken as the origin.

A plane wave arriving from direction $\bs{s}$ produces complex sound pressures
\begin{equation}
p_{+} = P_0 
e^{+ \ii k\bs{s}^{T}\bs{x}_{+}}, \;\;\
p_{-} = P_0 
e^{+ \ii k\bs{s}^{T}\bs{x}_{-}},
\end{equation}
at the two microphone positions.

The outputs of the microphones are expressed as
\begin{equation}
Y_{+}
=
D_{+}p_{+},
\;\;\;
Y_{-}
=
D_{-}p_{-},
\label{eq:plane_outputs}
\end{equation}
where $D_{+}$ and $D_{-}$ denote the directivities of the forward- and rear-facing cardioid microphones, respectively.

For ideal cardioid directivity,
\begin{equation}
D_{+} = \frac{1}{2}
\left(
1+\bs{s}^{T}\bs{r}
\right), \;\;
D_{-} = \frac{1}{2}
\left(
1-\bs{s}^{T}\bs{r}
\right).
\end{equation}
Then, introducing
\begin{equation}
\mu =
\cos\theta,
\;\;\;
\phi =
k\frac{d}{2}
\bs{s}^{T}\bs{r},
\label{eq:mu_phi}
\end{equation}
the microphone outputs become
\begin{equation}
Y_{+} = \frac{P_0}{2}
(1+\mu) e^{+ \ii \phi}, \;\;
Y_{-} = \frac{P_0}{2}
(1-\mu) e^{- \ii \phi}.
\label{eq:Ypm}
\end{equation}

The C-C method estimates pressure and particle velocity from the sum and difference of the two microphone signals as
\begin{equation}
p_{\mathrm{est}} =
Y_{+}+Y_{-}, \;\;
u_{\mathrm{est}} = -
\frac{Y_{+}-Y_{-}}{Z_0},
\label{eq:p_andu_est}
\end{equation}
where $Z_0$ denotes the real-valued characteristic acoustic impedance of the homogeneous, lossless medium (mass density times speed of sound). It is used as the reference impedance in the microphone-pair reconstruction.

Throughout this paper, the quantities obtained from the sum and difference signals are denoted as estimated pressure, estimated particle velocity, and estimated acoustic intensity. 
These quantities should be interpreted as estimates derived from the C-C operator rather than exact physical quantities, since the sum and difference responses of two cardioid microphones cannot realize perfectly omnidirectional and dipole directivities.

The estimated acoustic intensity is therefore
\begin{equation}
I_{\mathrm{est}}
=
\frac{1}{2}
\Re
\left[
p_{\mathrm{est}}
u_{\mathrm{est}}^{*}
\right],
\end{equation}
where $\Re$ indicates the real part. Substituting Eq.~(\ref{eq:Ypm}) yields
\begin{equation}
I_{\mathrm{est}}
= -
\frac{|P_0|^2}{2Z_0}
\left(
|D_{+}|^2
-
|D_{-}|^2
\right).
\label{eq:Iest_plane}
\end{equation}
Interestingly, the propagation phase difference caused by the finite microphone spacing completely vanishes in Eq.~(\ref{eq:Iest_plane}).
Consequently, the directional intensity estimate depends only on the microphone directivities and is independent of the spacing $d$.

Note that the cancellation of the spacing-dependent phase term is valid for the active intensity considered in this study, which is the quantity most frequently used in practical acoustic measurements. Reactive intensity components may still exhibit a dependence on microphone spacing.

This observation explains why the C-C method is fundamentally less sensitive to microphone spacing than conventional pressure-gradient approaches.
However, any deviation from the ideal cardioid response directly affects the estimated pressure, particle velocity, and intensity.

\subsection{Spherical-Wave Formulation}

Next, consider a spherical wave radiated from a point source located at a finite distance.
Let the source position be denoted by $\bs{x}_{\mathrm{s}}$.
The complex sound pressure at an arbitrary microphone position $\bs{x}$ is expressed as
\begin{equation}
p(\bs{x}) = A
\frac{e^{-\ii kR}}{R},
\label{eq:spherical_pressure}
\end{equation}
where
\begin{equation}
R =
\left\|
\bs{x} - \bs{x}_{\mathrm{s}}
\right\|
\end{equation}
is the distance between the source and the microphone, and $A$ is a complex amplitude related to the source strength.

For the two microphones located at $\bs{x}_{+}$ and $\bs{x}_{-}$, the propagation distances are
\begin{equation}
R_{+} =
\left\|
\bs{x}_{+}
-
\bs{x}_{\mathrm{s}}
\right\|,
\;\;
R_{-}
=
\left\|
\bs{x}_{-}
-
\bs{x}_{\mathrm{s}}
\right\|.
\end{equation}

The corresponding microphone outputs are therefore given by
\begin{equation}
Y_{+}
=
A D_{+}
\frac{e^{-\ii kR_{+}}}{R_{+}},
\;\;
Y_{-}
=
A D_{-}
\frac{e^{-\ii kR_{-}}}{R_{-}},
\label{eq:spherical_output}
\end{equation}
where $D_{+}$ and $D_{-}$ denote the complex directivities evaluated for the local incidence directions at the two microphone positions.

Unlike the plane-wave case, the local direction of arrival differ between the two microphones.
They are given by
\begin{equation}
\bs{s}_{+}
=
\frac{
\bs{x}_{\mathrm{s}} - \bs{x}_{+}
}{
R_{+}
},
\;\;
\bs{s}_{-}
=
\frac{
\bs{x}_{\mathrm{s}}
-
\bs{x}_{-}
}{
R_{-}
}.
\end{equation}

Accordingly, the microphone directivities should be written more precisely as
\begin{equation}
D_{+}
=
D(\bs{r},\bs{s}_{+}),
\;\;
D_{-}
=
D(\bs{r},\bs{s}_{-}).
\end{equation}

As in the plane-wave formulation, the C-C method estimates pressure and the particle-velocity-equivalent component from the sum and difference signals:
\begin{equation}
p_{\mathrm{est}}
=
Y_{+}
+
Y_{-} 
=
A
\left (
D_{+}
\frac{e^{-\ii kR_{+}}}{R_{+}}
+
D_{-}
\frac{e^{-\ii kR_{-}}}{R_{-}}
\right ),
\label{eq:spherical_pressure_est}
\end{equation}
and
\begin{equation}
u_{\mathrm{est}}
= -
\frac{
Y_{+}
-
Y_{-}
}{
Z_0
} 
= -
\frac{A}{Z_0}
\left (
D_{+}
\frac{e^{-\ii kR_{+}}}{R_{+}}
-
D_{-}
\frac{e^{-\ii kR_{-}}}{R_{-}}
\right ).
\label{eq:spherical_velocity_est}
\end{equation}
Here, $Z_0$ is the real-valued characteristic acoustic impedance defined in Eq.~(\ref{eq:p_andu_est}).

The corresponding intensity estimate is
\begin{equation}
I_{\mathrm{est}}
= 
\frac{1}{2}
\Re
\left[
p_{\mathrm{est}}
u_{\mathrm{est}}^{*}
\right].
\end{equation}

Substitution of Eqs.~(\ref{eq:spherical_pressure_est}) and
(\ref{eq:spherical_velocity_est}) gives
\begin{equation}
I_{\mathrm{est}}
= -
\frac{|A|^2}{2Z_0}
\left(
\frac{|D_{+}|^2}{R_{+}^{2}}
-
\frac{|D_{-}|^2}{R_{-}^{2}}
\right).
\label{eq:spherical_intensity_est}
\end{equation}

Thus, as in the plane-wave case, the cross terms associated with the two microphone signals cancel in the active-intensity calculation.
However, in the spherical-wave case, the two microphones generally experience different propagation distances, amplitudes, and local incidence directions.

When the source distance is sufficiently large compared with the microphone spacing, the approximations
\begin{equation}
R_{+}
\simeq
R_{-}
\simeq
R
\end{equation}
may be applied.
Equation~(\ref{eq:spherical_intensity_est}) then reduces to
\begin{equation}
I_{\mathrm{est}}
\simeq -
\frac{|A|^2}{2Z_0R^2}
\left(
|D_{+}|^2
-
|D_{-}|^2
\right).
\label{eq:spherical_intensity_far}
\end{equation}
Therefore, even for spherical-wave incidence, the explicit dependence on the microphone spacing disappears under the far-field approximation.
Nevertheless, because the wavefronts arriving at the two microphones are not exactly identical, spherical-wave incidence introduces additional error factors compared with the plane-wave case, including differences in propagation distance, amplitude, and local incidence direction.


\section{Directional Intensity Observation and Reconstruction}

Using the procedure described above, directional intensity can be estimated along an arbitrary axis $\bs{r}$ for both plane-wave and spherical-wave sound fields.
The resulting quantity may be interpreted as the projection of the three-dimensional acoustic intensity vector onto the microphone-pair axis.

\subsection{Reconstruction of Three-Dimensional Intensity from Directional Measurements}

Let the acoustic intensity vector at a given point be
\begin{equation}
\bs{I}
=
\begin{bmatrix}
I_x \;\; &
I_y \;\; &
I_z
\end{bmatrix}^{T}.
\end{equation}
The unit vector corresponding to the axis of the $i$th microphone pair is denoted by
\begin{equation}
\bs{r}_i
=
\begin{bmatrix}
r_{i,x} \;\; &
r_{i,y} \;\; &
r_{i,z}
\end{bmatrix}^{T}.
\end{equation}

The directional intensity measured along this axis is given by
\begin{equation}
I_{r,i}
=
\bs{r}_i^{T}
\bs{I}.
\label{eq:direction_projection}
\end{equation}
Equation (\ref{eq:direction_projection}) indicates that each measurement corresponds to a projection of the three-dimensional intensity vector onto the observation direction.

When measurements are performed along $N_r$ directions, the directional intensities can be collected into the vector
\begin{equation}
\bs{I}_{r}
=
\begin{bmatrix}
I_{r,1} &
I_{r,2} &
\cdots &
I_{r,N_r}
\end{bmatrix}^{T}.
\end{equation}
Defining the observation matrix
\begin{equation}
\bs{R}
=
\begin{bmatrix}
\bs{r}_1 \;\; & \bs{r}_2 \;\; & \cdots &  \bs{r}_{N_r}
\end{bmatrix}^T,
\label{eq:R}
\end{equation}
the directional measurements may be written compactly as
\begin{equation}
\bs{I}_{r}
=
\bs{R}
\bs{I}.
\label{eq:forward_projection}
\end{equation}

Equation (\ref{eq:forward_projection}) shows that the directional intensities are obtained by projecting the acoustic intensity vector onto a set of observation directions.
Consequently, the original three-dimensional intensity vector can be reconstructed from multiple directional measurements.

For $N_r > 3$, the reconstruction problem becomes overdetermined.
The least-squares estimate of the intensity vector, $\widehat{\bs{I}}$ is therefore given by
\begin{equation}
\widehat{\bs{I}}
=
\left(
\bs{R}^{T}
\bs{R}
\right)^{-1}
\bs{R}^{T}
\bs{I}_{r}.
\label{eq:least_square}
\end{equation}


\subsection{Tight-Frame Reconstruction}

Equation (\ref{eq:least_square}) provides the general least-squares solution for reconstructing the acoustic intensity vector from multiple directional measurements.
This expression becomes considerably simpler when the observation directions are chosen appropriately.

In the present study, attention is focused on a class of directional configurations known as tight frames.
A set of unit vectors
$\{\bs{r}_i\}_{i=1}^{N}$
forms a tight frame if
\begin{equation}
\sum_{i=1}^{N_r}
\bs{r}_i
\bs{r}_i^{T} = F \mathbf{I},
\label{eq:tightframe}
\end{equation}
where $F$ is the frame constant and $\mathbf{I}$ denotes the $3\times3$ identity matrix \cite{Casazza2006,Barg2015},\cite[p.348]{Waldron2018},\cite{Hughes2021}.

From Eq.~(\ref{eq:R}), it follows that
\begin{equation}
\bs{R}^{T}
\bs{R}
=
\sum_{i=1}^{N_r}
\bs{r}_i
\bs{r}_i^{T}
=
F\mathbf{I}.
\end{equation}
Substituting this relation into Eq.~(\ref{eq:least_square}) yields
\begin{equation}
\widehat{\bs{I}}
=
\frac{1}{F}
\bs{R}^{T}
\bs{I}_{r}
=
\frac{1}{F}
\sum_{i=1}^{N_r}
I_{r,i}
\bs{r}_i.
\label{eq:tightframe_reconstruction}
\end{equation}
Equation (\ref{eq:tightframe_reconstruction}) shows that the reconstruction can be achieved simply by summing the directional intensity measurements weighted by their corresponding direction vectors.
No matrix inversion is required, and the resulting estimator possesses a high degree of symmetry.


\subsection{Tight-Frame Configurations Considered}

\begin{figure}[t]
\centering
\includegraphics[bb={0 0 254 283},width=0.8\linewidth,pagebox=artbox]{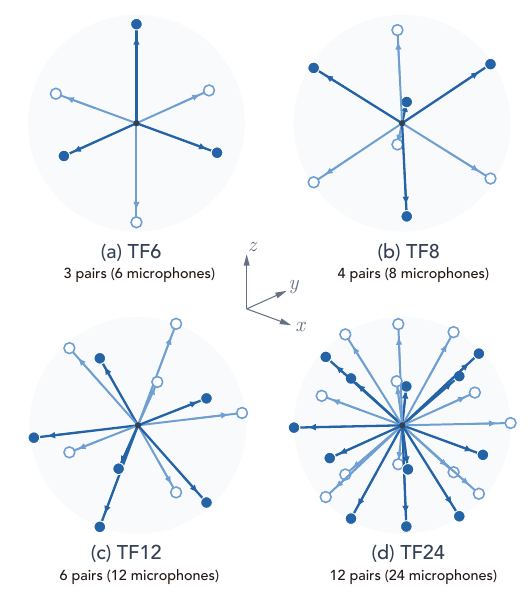}
\caption{Arrangement of tight-frame configurations assumed in this study. }
\label{fig:TFs}
\end{figure}

Four tight-frame microphone-array configurations were examined in this study:
\begin{itemize}
\item TF6: six microphone positions corresponding to three orthogonal axes ($x$, $y$, and $z$ directions),
\item TF8: eight microphone positions located at the vertices of a cube,
\item TF12: twelve microphone positions corresponding to the vertices of a regular icosahedron,
\item TF24: twenty-four microphone positions distributed over three elevation layers at $45^\circ$ intervals in azimuth.
\end{itemize}
These are illustrated in Fig. \ref{fig:TFs}.
For each configuration, opposing microphones form cardioid pairs, and the directional intensities obtained from the corresponding axes are combined according to Eq.~(\ref{eq:tightframe_reconstruction}).

The observation matrices used for the four configurations are given by
\[
\mathbf{R}_{\mathrm{TF6}} =
\begin{bmatrix}
1\;\;&0\;\;&0\\
0\;\;&1\;\;&0\\
0\;\;&0\;\;&1
\end{bmatrix}, \;\;
\mathbf{R}_{\mathrm{TF8}} =
\frac{1}{\sqrt{3}}
\begin{bmatrix}
1 \;\;& 1 \;\;& 1\\
1 \;\;& 1 \;\;&-1\\
1 \;\;&-1 \;\;& 1\\
1 \;\;&-1 \;\;&-1
\end{bmatrix}, 
\]
\[
\mathbf{R}_{\mathrm{TF12}}
=
\begin{bmatrix}
0 \;\;& 0.526 \;\;& 0.851\\
0 \;\;& -0.526 \;\;& 0.851\\
0.526 \;\;& 0.851 \;\;& 0\\
-0.526 \;\;& 0.851 \;\;& 0\\
0.851 \;\;& 0 \;\;& 0.526\\
-0.851 \;\;& 0 \;\;& 0.526
\end{bmatrix},
\]
\begin{equation}
\bs{R}_{\mathrm{TF24}}
=
\begin{bmatrix}
 1/ \sqrt{2} \;\;& 0                  \;\;& 1/ \sqrt{2} \\
1/2          \;\;& 1/2            \;\;& 1/ \sqrt{2} \\
 0                \;\;& 1/ \sqrt{2}   \;\;& 1/ \sqrt{2} \\
-1/2          \;\;& 1/2            \;\;& 1/ \sqrt{2} \\
-1/ \sqrt{2} \;\;& 0                  \;\;& 1/ \sqrt{2} \\
-1/2          \;\;&-1/2            \;\;& 1/ \sqrt{2} \\
 0                \;\;&-1/ \sqrt{2}   \;\;& 1 / \sqrt{2} \\
 1/2          \;\;&-1/2            \;\;& 1 / \sqrt{2} \\
 1                \;\;& 0                  \;\;& 0 \\
 1/ \sqrt{2} & 1/ \sqrt{2}   \;\;& 0 \\
 0                \;\;& 1                  \;\;& 0 \\
-1/ \sqrt{2} \;\;& 1 / \sqrt{2}   \;\;& 0
\end{bmatrix}.
\label{eq:TF24_R}
\end{equation}
The corresponding frame constants are
\begin{equation}
F=
\left\{
1,\,
\frac{4}{3},\,
2,\,
4
\right\}
\end{equation}
for TF6, TF8, TF12, and TF24, respectively.

The TF12 configuration consists of six antipodal axes passing through the twelve vertices of a regular icosahedron.
The vertex coordinates are obtained by normalizing
\[
(0,\pm1,\pm\varphi),
\quad
(\pm1,\pm\varphi,0),
\quad
(\pm\varphi,0,\pm1),
\]
where
\begin{equation}
\varphi=\frac{1+\sqrt{5}}{2}
\end{equation}
is the golden ratio.
The twelve vertices of a regular icosahedron form a spherical 5-design and therefore constitute a finite unit-norm tight frame in $\mathbb{R}^{3}$
\cite{Conway1996,Waldron2018,Hughes2021}.
This high degree of symmetry is expected to provide strong suppression of direction-dependent measurement errors.

The TF24 configuration was originally introduced by the author for sound-field recording and reproduction \cite{Omoto2020} and was later applied to directional diffuseness estimation, where it exhibited particularly robust directional averaging properties \cite{Omoto2026}.
The present study investigates the origin of this robustness from the viewpoint of sound-intensity reconstruction.


\section{Error Analysis}

\subsection{Error Propagation in Tight-Frame Reconstruction}

When the microphone spacing approaches zero and the microphone directivity is perfectly cardioid, the sum and difference signals of a microphone pair provide ideal estimates of sound pressure and particle velocity.
In practice, however, the microphone spacing is finite, the directivity deviates from the ideal cardioid response, microphone sensitivities may be mismatched, and additive noise is inevitably present.
Consequently, the estimated directional intensity contains errors.
Hereafter, the microphone directivity is treated as frequency dependent. The angular frequency is denoted by $\omega=2\pi f$, where $f$ denotes a frequency.

To describe these effects of errors in a unified manner, the estimated directional intensity is modeled as
\begin{equation}
I_{r,i}^{\mathrm{est}}
=
\left[
\alpha(\omega)
+
\delta_i(\omega)
\right]
I_{r,i}
+
\epsilon_i ,
\label{eq:err_model}
\end{equation}
where $I_{r,i}^{\mathrm{est}}$ denotes the estimated directional intensity along the $i$th observation axis and $\alpha(\omega)$ represents a direction-independent average transfer coefficient common to all observation directions, whereas $\delta_i(\omega)$ represents a direction-dependent error component specific to the $i$th observation direction. 
Here, $\epsilon_i$ denotes additive measurement noise in the $i$th direction.
Here, $I_{r,i}$ is the corresponding true directional intensity given by
\begin{equation}
I_{r,i}
=
\bs{r}_i^{T}
\bs{I} .
\end{equation}

The coefficient $\alpha(\omega)+\delta_i(\omega)$ does not represent the microphone frequency response alone.
Rather, it describes the cumulative error introduced by the entire C-C operator, including sound reception by the non-ideal directional microphones, pressure estimation, particle-velocity estimation, and the subsequent intensity calculation.

Using the tight-frame reconstruction of Eq.~(\ref{eq:tightframe_reconstruction}), the reconstructed three-dimensional intensity vector becomes
\begin{equation}
\widehat{\bs I}
=
\frac{1}{F}
\sum_{i=1}^{N_r}
I_{r,i}^{\mathrm{est}}
\bs r_i .
\label{eq:TF24_reconstruction}
\end{equation}
Substituting Eq.~(\ref{eq:err_model}) into Eq.~(\ref{eq:TF24_reconstruction}) and using the relation 
\[
I_{r,i}=\bs{r}_i^T\bs{I},
\;\; I_{r,i}\bs{r}_i = (\bs{r}_i^T \bs{I}) \bs{r}_i  =   \bs{r}_i\bs{r}_i^T\bs{I}
\]
yields
\begin{align}
\widehat{\bs I}
& = \frac{1}{F} \sum_{i=1}^{N_r}
\left[ \left\{ \alpha(\omega) + \delta_i(\omega) \right\} I_{r,i} + \epsilon_i \right] \bs{r}_i \notag \\
& = \frac{\alpha(\omega)}{F} \sum_{i=1}^{N_r} \bs{r}_i \bs{r}_i^{T} \bs{I} + \frac{1}{F} \sum_{i=1}^{N_r} \delta_i(\omega) \bs{r}_i \bs{r}_i^{T} \bs{I}
+
\frac{1}{F} \sum_{i=1}^{N_r} \epsilon_i \bs{r}_i
\notag \\
&= \alpha(\omega) \bs{I} +
\frac{1}{F}
\sum_{i=1}^{N_r}
\delta_i(\omega)
\bs{r}_i
\bs{r}_i^{T}
\bs{I}
+
\frac{1}{F}
\sum_{i=1}^{N_r}
\epsilon_i
\bs{r}_i .
\label{eq:TF24_expand2}
\end{align}
The first term represents a global gain factor and therefore does not alter the direction of the reconstructed intensity vector.
The second term represents the propagation of direction-dependent errors through the tight-frame reconstruction process.
The third term corresponds to additive measurement noise.

Consequently, the effectiveness of a tight-frame configuration is primarily determined by the degree to which the second term is suppressed.
The following analysis therefore focuses on the behavior of the direction-dependent error component.


\subsection{Spherical-Harmonic Representation of Direction-Dependent Errors}

The discrete coefficients
$\delta_i(\omega)$
may be regarded as samples of a continuous directional-error function
$\delta(\bs{r}, \omega)$
defined on the unit sphere:
\begin{equation}
\delta_i(\omega)
=
\delta(\bs{r}_i,\omega).
\end{equation}
Using this notation, the direction-dependent error matrix appearing in Eq.~(\ref{eq:TF24_expand2}) can be written as
\begin{equation}
\bs{E}
=
\frac{1}{F}
\sum_{i=1}^{N_r}
\delta(\bs{r}_i,\omega)
\,
\bs{r}_i
\bs{r}_i^{T}.
\label{eq:error_matrix_continuous_sample}
\end{equation}
Since
$\delta(\bs{r},\omega)$
is a continuous function defined on the sphere $S^2$,
it can be expanded in terms of spherical harmonics as
\begin{equation}
\delta(\bs{r},\omega)
=
\sum_{n=0}^{\infty}
\sum_{m=-n}^{n}
a_n^m(\omega)
Y_n^m(\bs{r}),
\label{eq:delta_SH}
\end{equation}
where $Y_n^m(\bs{r})$
denotes the spherical harmonic of degree $n$ and order $m$, and $a_n^m(\omega)$ is the corresponding expansion coefficient.
We adopt the standard mathematical nomenclature here; in some acoustics literature, $n$ and $m$ are instead referred to as the order and degree, respectively \cite{Rafaely2015}.

Substituting Eq.~(\ref{eq:delta_SH}) into Eq.~(\ref{eq:error_matrix_continuous_sample}) yields
\begin{equation}
\bs{E}
=
\frac{1}{F}
\sum_{i=1}^{N_r}
\left[
\sum_{n=0}^{\infty}
\sum_{m=-n}^{n}
a_n^m(\omega)
Y_n^m(\bs{r}_i)
\right]
\bs{r}_i
\bs{r}_i^{T}.
\end{equation}
Interchanging the order of summation gives
\begin{equation}
\bs{E}
=
\sum_{n=0}^{\infty}
\sum_{m=-n}^{n}
a_n^m(\omega)
\bs{E}_n^m ,
\label{eq:error_decomposition}
\end{equation}
where
\begin{equation}
\bs{E}_n^m
=
\frac{1}{F}
\sum_{i=1}^{N_r}
Y_n^m(\bs{r}_i)
\,
\bs{r}_i
\bs{r}_i^{T}.
\label{eq:Enm}
\end{equation}
Equation~(\ref{eq:Enm}) describes how a spherical-harmonic error component of degree $n$ and order $m$ propagates through the reconstruction process associated with a given microphone configuration.

Consequently, the directional-error suppression capability of a tight-frame configuration can be understood by examining the magnitude of
$\bs{E}_n^m$.
If
$\bs{E}_n^m$
is small, the corresponding spherical-harmonic error component is effectively suppressed after reconstruction.
Conversely, large values of
$\bs{E}_n^m$
indicate that the corresponding directional-error component is transmitted to the reconstructed intensity vector.

To evaluate the suppression capability of a tight-frame arrangement, the energy over all orders is combined into the following quantity:
\begin{equation}
G_n
=
\left(
\sum_{m=-n}^{n}
\|
\bs E_n^m
\|_F^2
\right)^{1/2},
\label{eq:G_n}
\end{equation}
where
$\|\cdot\|_F$
denotes the Frobenius norm.

\begin{figure}[t]
\centering
\includegraphics[bb={0 0 450 327},width=\linewidth,pagebox=artbox]{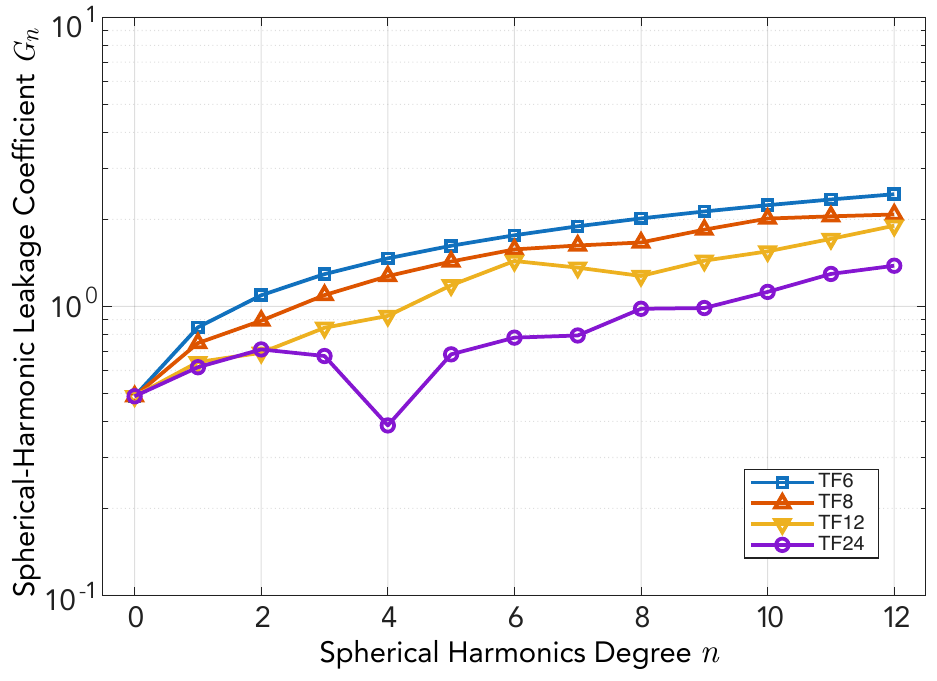}
\caption{Spherical-harmonic leakage coefficient $G_n$ associated with each tight-frame configuration.
$G_n$ quantifies the degree to which an error component of spherical-harmonic degree $n$ is transmitted to the reconstructed intensity vector.
Smaller values correspond to stronger error suppression.}
\label{fig:G_n}
\end{figure}

The quantity $G_n$ represents the leakage of a degree-$n$ direction-dependent error through the reconstruction process.
A small value of $G_n$ indicates that spherical-harmonic errors of degree $n$ are effectively suppressed by the microphone geometry,
whereas large values imply that such errors remain in the reconstructed intensity vector.
Figure~\ref{fig:G_n} shows the values of $G_n$ calculated up to $n=12$ for TF6, TF8, TF12, and TF24.

Among the four configurations, TF24 exhibits the smallest values over almost all degrees, indicating the strongest suppression of direction-dependent errors.

\subsection{Modeling of Practical Microphone Directivity Errors}

As a principal source of the direction-dependent error $\delta(\bs{r},\omega)$, the directivity error of a commercially available cardioid microphone is considered.

Assuming axisymmetric directivity, the spherical-harmonic expansion reduces to the $m=0$ terms only and can therefore be represented by a Legendre-polynomial expansion. In this case, the directivity depends only on the polar angle $\theta$.

In the present study, the directivity of a DPA2012 cardioid microphone was measured in an anechoic chamber.
Impulse responses were acquired at a source distance of 2.5~m while the microphone was rotated in increments of $3^\circ$.

From the measured impulse responses, the complex directivity response $D_{\mathrm{meas}}(\theta,\omega)$ was evaluated at octave-spaced center frequencies from 63~Hz to 16~kHz. 
At each center frequency, the representative magnitude was obtained by averaging the FFT magnitudes within a one-third-octave bandwidth, whereas the phase was taken from the FFT bin nearest to the center frequency.

A small frequency-dependent offset of the maximum-sensitivity direction was regarded as an error in the measured angular origin.
Accordingly, for each center frequency, the measured response was rotated so that its maximum-magnitude direction coincided with $\theta=0$. 
The complex response was then normalized by its on-axis value. 
Under the assumption of axisymmetric directivity, the aligned response over $0\leq\theta\leq\pi$ was used to construct the directivity model.

The deviation of the measured directivity from the ideal cardioid response, $\delta_\mathrm{meas}$, is then defined as
\begin{equation}
\delta_\text{meas} (\theta,\omega)
=
D_{\mathrm{meas}}(\theta,\omega)
-
\frac{1+\cos\theta}{2}.
\end{equation}

The resulting error function was expanded using Legendre polynomials as
\begin{equation}
\delta_\text{meas}(\theta,\omega)
=
\sum_{n=0}^{N_L}
\Delta c_n(\omega)
P_n(\cos\theta),
\label{eq:expanded_error}
\end{equation}
where
$\Delta c_n(\omega)$
denotes the Legendre-expansion coefficient associated with degree $n$.
The complex Legendre coefficients were determined by an unweighted least-squares fit to the measured responses sampled uniformly in $\theta$. 
In this study, the expansion was truncated at $N_L=8$.

\begin{figure}[t]
\centering
\includegraphics[bb={0 0 510 413},width=\linewidth,pagebox=artbox]{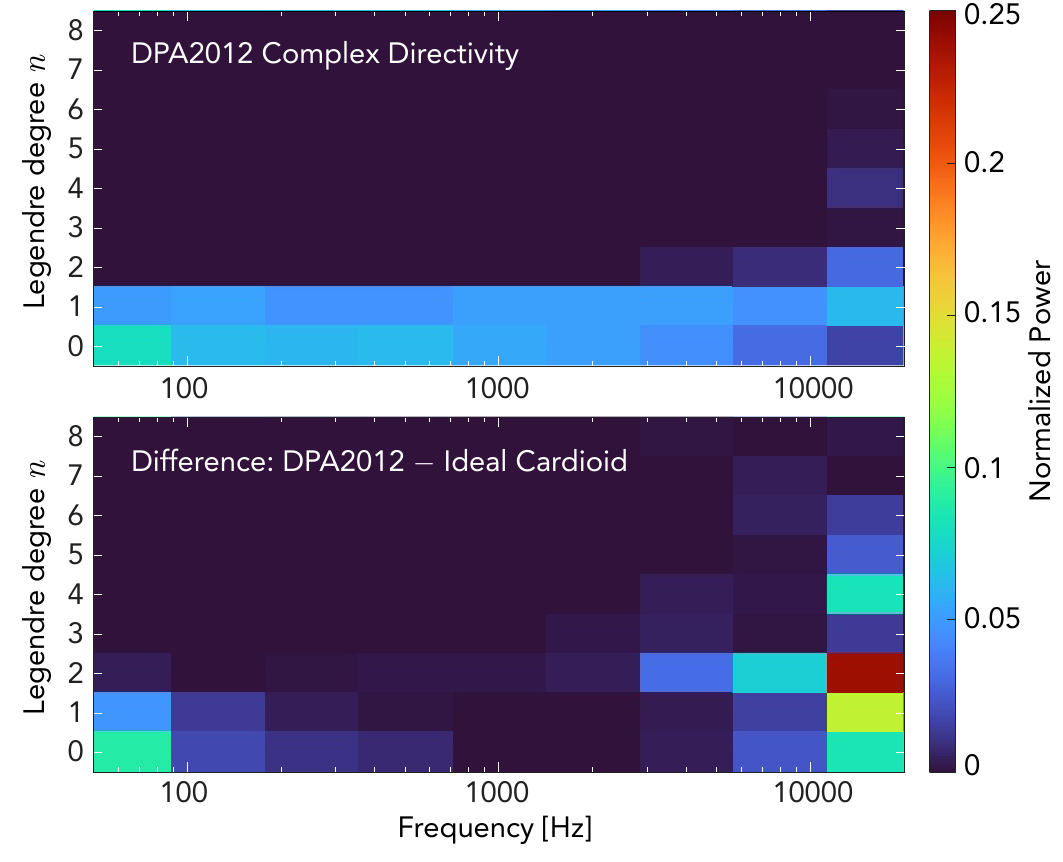}
\caption{Normalized coefficient-energy distributions of the measured DPA2012 directivity (top) and its deviation from the ideal cardioid response (bottom). 
Each distribution was normalized by its total coefficient energy summed over all center frequencies and Legendre degrees.}
\label{fig:c_n}
\end{figure}

Figure~\ref{fig:c_n} shows the normalized coefficient-energy distributions of the measured directivity and its deviation from the ideal cardioid response. 
For each panel, the squared coefficient magnitudes were normalized by the corresponding total coefficient energy summed over all center frequencies and Legendre degrees.

The upper panel of Fig.~\ref{fig:c_n} shows the Legendre expansion of the measured microphone directivity itself.
As expected for a cardioid microphone, the directivity is described predominantly by the zeroth-degree (omnidirectional) and first-degree (dipole) components.
Higher-order contributions become slightly more noticeable at higher frequencies.

The lower panel of Fig.~\ref{fig:c_n} shows the deviation from the ideal cardioid response. 
A relatively strong zeroth-degree component is observed at 63~Hz, while second-degree components become increasingly important at higher frequencies.

Overall, the majority of the energy remains concentrated in the low-degree terms.
The directivity error of the DPA2012 microphone can therefore be characterized primarily by a small number of low-order angular components.

\subsection{Suppression of Directivity Errors by Tight Frames}

The analysis presented above provides two complementary pieces of information.

First, the coefficients $\Delta c_n(\omega)$ describe the angular structure of the microphone directivity error.
Second, the quantities $G_n$ describe how strongly directional errors of degree $n$ are transmitted through a given microphone geometry.

It is therefore natural to consider the combined effect of these two factors.
A microphone may exhibit large directivity errors at a certain degree, but such errors will have little influence on the reconstructed intensity vector if the corresponding value of $G_n$ is sufficiently small.
Conversely, even a relatively small directivity error may become important if the geometry exhibits a large leakage coefficient for the same degree.

To visualize this interaction, the following effective leakage indicator is introduced:
\begin{equation}
\Lambda(\omega)
=
\left(
\sum_{n=0}^{N_L}
|\Delta c_n(\omega)|^2
G_n^2
\right)^{1/2},
\label{eq:S}
\end{equation}
where $n$ denotes the Legendre degree.

Here,
$\Delta c_n(\omega)$
represents the Legendre-expansion coefficient of the microphone directivity error, and $G_n$ represents the geometry-dependent leakage coefficient introduced in
Eq.~(\ref{eq:G_n}).

Note that Eq.~(\ref{eq:S}) is not derived as an exact expression for the intensity-estimation error.
Rather, it serves as a physically interpretable metric that combines the magnitude of the microphone directivity error and the suppression capability of the microphone geometry.
In this sense, $\Lambda(\omega)$ may be regarded as an estimate of how efficiently directivity errors are transmitted into the reconstructed intensity vector.

\begin{figure}[t]
\centering
\includegraphics[bb={0 0 435 326},width=\linewidth,pagebox=artbox]{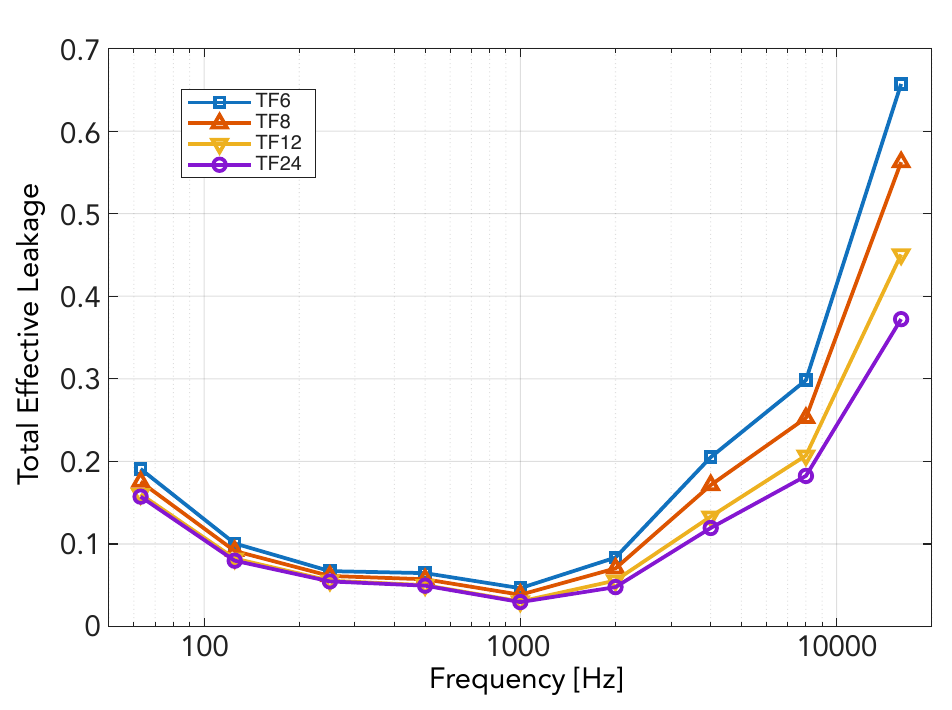}
\caption{Total effective leakage $\Lambda(\omega)$ for the DPA2012 microphone with several tight-frame configurations.
Smaller values indicate a lower susceptibility to direction-dependent errors, although $\Lambda(\omega)$
should be interpreted as a relative indicator rather than a direct measure of estimation error.}
\label{fig:Lambda}
\end{figure}

Figure~\ref{fig:Lambda} shows the calculated values of $\Lambda(\omega)$ for the DPA2012 microphone combined with the four tight-frame configurations considered in this study.

As expected from the behavior of $G_n$, the effective leakage decreases systematically as the number of observation directions increases.
Among the investigated configurations, TF24 exhibits the smallest values over the entire frequency range, indicating the strongest suppression of directivity-related errors.


\section{Numerical Simulations}

To evaluate the practical performance of the proposed approach, numerical simulations were conducted under relatively demanding conditions, including spherical-wave propagation and additive noise.

The estimation procedure described in Section 2 was used throughout all simulations. Two different microphone directivity models were considered: an ideal cardioid model and the measured DPA2012 directivity model.

The objective was to investigate the accuracy of intensity estimation obtained by combining the C-C method with various tight-frame microphone configurations and to examine how microphone directivity errors influence the reconstructed intensity vector.

The simulation conditions were as follows:
\begin{itemize}
\item Microphone configurations:
TF6, TF8, TF12, and TF24.
\item Microphone directivity:
ideal cardioid / measured DPA2012 directivity model.
In both cases, the same C-C processing equations described in Section 2 was used. Only the microphone directivity model employed to generate the simulated microphone signals was changed.

Frequency-dependent scaling coefficients were applied so that the amplitudes of the sum and difference signals matched the corresponding on-axis responses. 
These coefficients are analogous to the $\alpha$ and $\beta$ parameters introduced by Hanyu and Hoshi~\cite{Hanyu2026}.
\item Source directions:
3000 directions uniformly distributed over the sphere using a
Fibonacci-sphere sampling scheme \cite{Gonzalez2010}.
\item Source distance:
spherical-wave sources located at
$2.5~\mathrm{m}\pm20\%$
from the array center.
\item Frequency:
For each octave band from 63 Hz to 16 kHz, 50 frequencies were randomly sampled from a uniform distribution on the logarithmic frequency axis.
Combined with the 3000 source directions generated by the Fibonacci-sphere sampling scheme, this resulted in 150,000 source-frequency realizations per octave band.
%
\item Microphone spacing:
0.02, 0.05, 0.10, and 0.20 m.
\item Additive noise:
signal-to-noise ratios of
$\infty$, 40 dB, and 20 dB, corresponding to noise levels of $-\infty$, $-40$, and $-20$ dB relative to the signal RMS.
\item Error metrics:
The direction error $\varepsilon_{\mathrm{dir}}$ and the signed
intensity-level error $\varepsilon_{\mathrm{L}}$ were defined as
\[
\varepsilon_{\mathrm{dir}}
=
\cos^{-1}
\left(
\bs{s}^{\mathrm{T}}\widehat{\bs{s}}
\right),
\qquad
\widehat{\bs{s}}
=
-\widehat{\bs{I}} /   \left\| \widehat{\bs{I}} \right\|,
\]
where $\widehat{\bs{s}}$ is the estimated unit vector pointing toward
the source and $\bs{s}$ is the true unit vector pointing from the array
center toward the source. The minus sign accounts for the fact that the
active-intensity vector points in the propagation direction, opposite
to the source direction. The direction error is expressed in degrees.

The signed intensity-level error was defined as
\[
\varepsilon_{\mathrm{L}}
=
10\log_{10}
\left(
\left\|\widehat{\bs{I}}\right\| / 
I_{\mathrm{ref}}
\right)
\quad \mathrm{dB},
\]
where $I_{\mathrm{ref}}$ is the analytical active-intensity magnitude
at the array center for the simulated spherical wave. Under the
normalization adopted here, $I_{\mathrm{ref}}=1/R^{2}$, where $R$ is
the source distance. A positive value indicates overestimation of the
intensity magnitude, whereas a negative value indicates
underestimation. The reported results are the medians of these errors
over all source-frequency realizations within each octave band.

\end{itemize}

\begin{figure*}[t]
\centering
\includegraphics[bb={0 0 559 611},width=0.9\textwidth,pagebox=artbox]{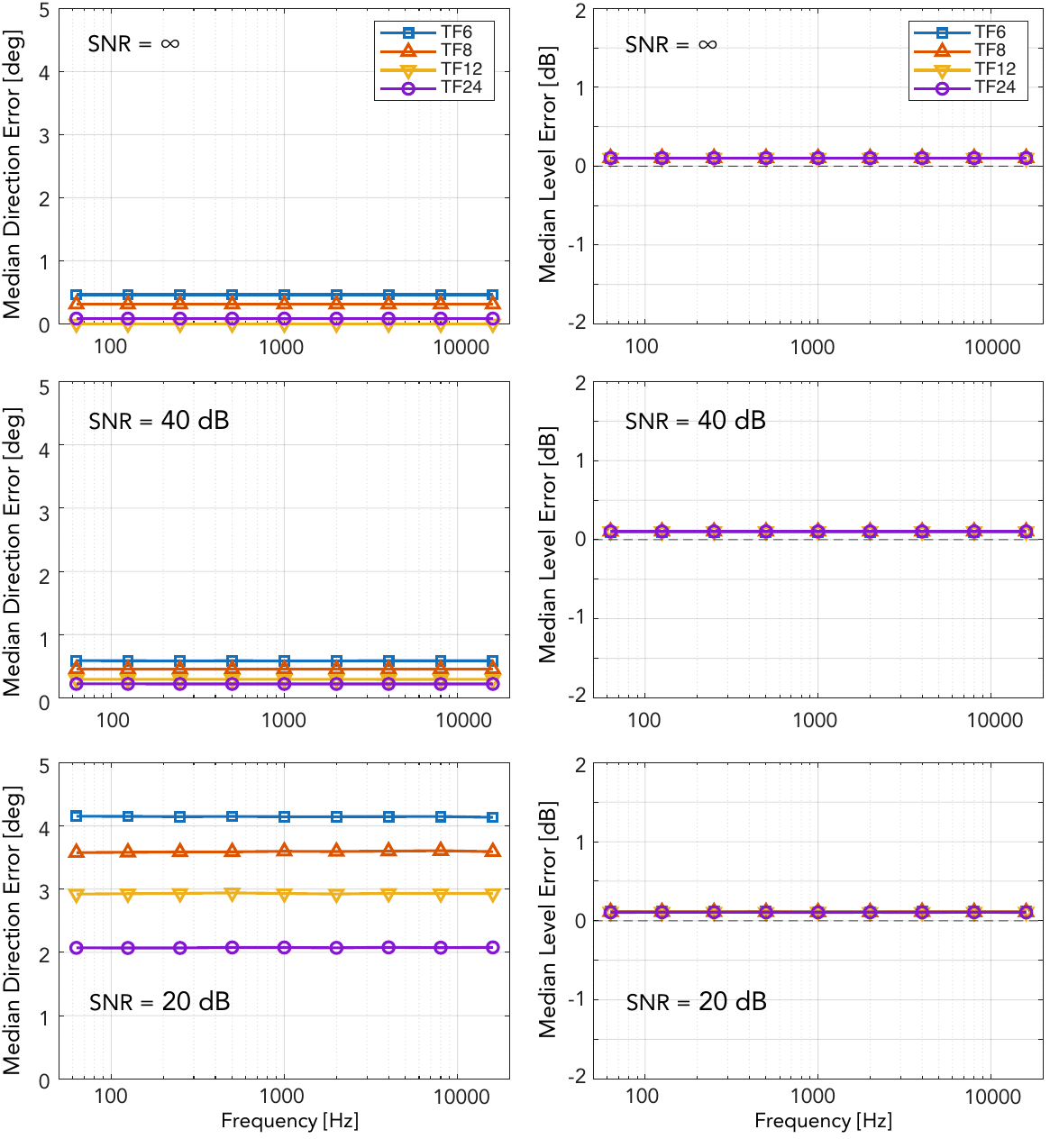}
\caption{Simulated intensity-estimation performance for spherical-wave incidence using ideal cardioid microphones.
Left: median direction-estimation error. Right: median signed intensity-level error.}
\label{fig:cardioid}
\end{figure*}

\begin{figure*}[t]
\centering
\includegraphics[bb={0 0 556 625},width=0.9\textwidth,pagebox=artbox]{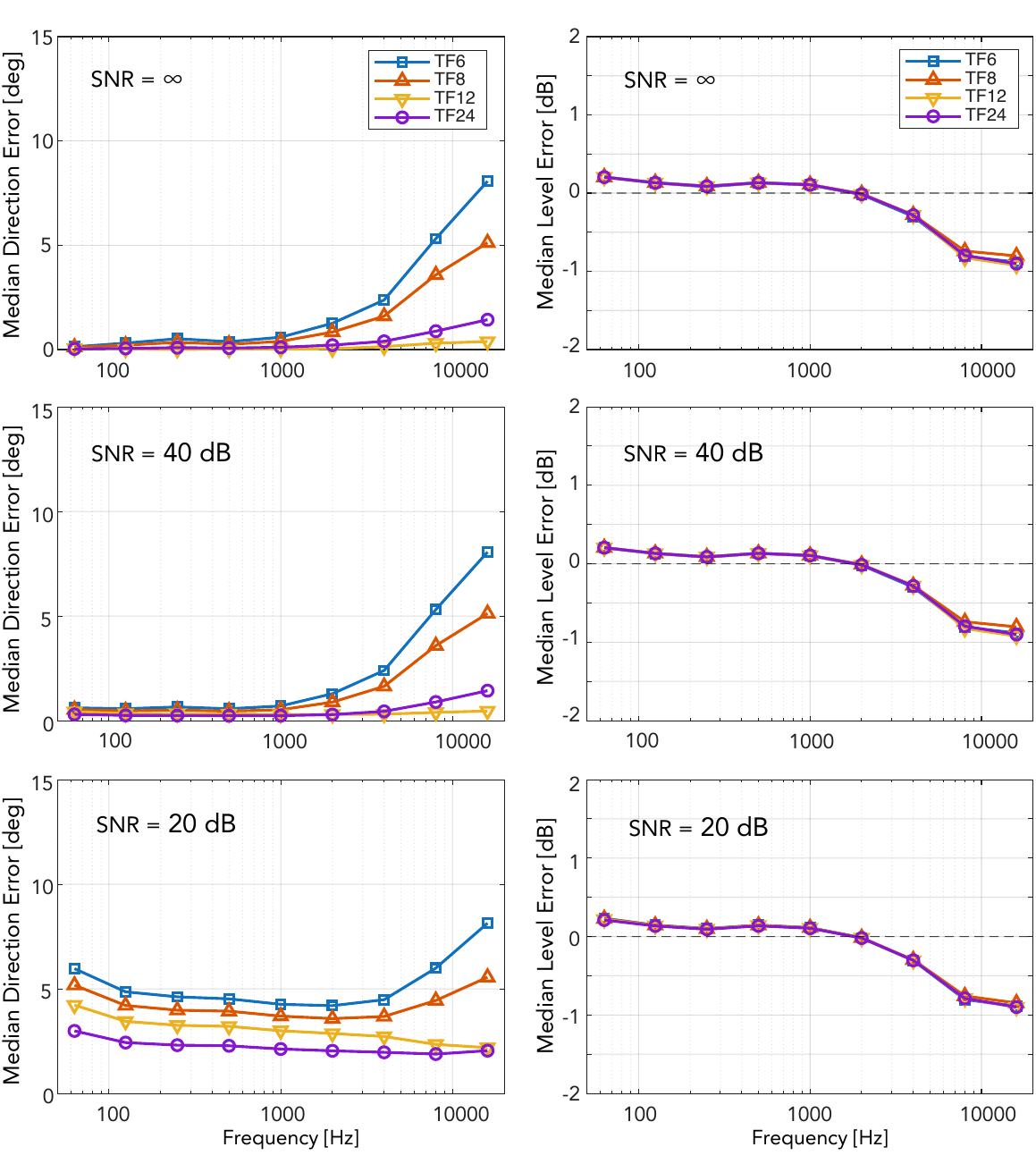}
\caption{Simulated intensity-estimation performance for spherical-wave incidence using the measured DPA2012 directivity model.
Left: median direction-estimation error. Right: median signed intensity-level error.}
\label{fig:DPA}
\end{figure*}

In the following, representative results obtained with a microphone spacing of 0.10 m are presented.
Figure~\ref{fig:cardioid} shows the results obtained with ideal cardioid microphones,
whereas Fig.~\ref{fig:DPA} shows the corresponding results obtained using the measured DPA2012 directivity model.

Figure~\ref{fig:cardioid} shows the performance obtained for ideal cardioid microphones.
As expected, the estimation accuracy is essentially independent of frequency because the microphone directivity perfectly satisfies the assumed model.

In the absence of additive noise, the median direction-estimation error is approximately 0.5$^\circ$ even for TF6.
The error decreases further as the number of observation directions increases.

At an SNR of 20 dB, the median direction-estimation errors remain within approximately 2--4.5$^\circ$ and decrease systematically as the number of observation directions increases.

The intensity-level estimation exhibits little dependence on the noise level. 
Errors remain within approximately 0.2 dB for all configurations.

In contrast, the influence of microphone geometry becomes much more apparent when the measured DPA2012 directivity is used (Fig.~\ref{fig:DPA}).

Even in the absence of additive noise, TF6 exhibits direction-estimation errors approaching 8$^\circ$ at 16 kHz.
At high SNR, TF12 and TF24 both yield small errors, and their numerical ordering is not consistent across noise conditions. 

The overall tendency is consistent with the behavior predicted by the effective leakage indicator $\Lambda(\omega)$ shown in Fig.~\ref{fig:Lambda}.
Configurations exhibiting smaller leakage coefficients also exhibit smaller direction-estimation errors.


An interesting observation is that TF24 exhibited the smallest direction-estimation error under the most severe noise condition.
This tendency is not fully explained by the leakage indicator $\Lambda(\omega)$, which primarily characterizes the suppression of direction-dependent directivity errors.

A possible explanation is that the larger number of microphone pairs in TF24 provides additional spatial redundancy.
Because the additive noise components are assumed to be mutually uncorrelated among channels, the reconstruction process effectively averages these perturbations over a greater number of measurement directions.
Consequently, the influence of random noise may be reduced more efficiently in TF24 than in the lower-order configurations.

The intensity-magnitude estimates exhibit substantially less sensitivity to microphone directivity errors than the direction estimates.
For all configurations, the estimated intensity tends to be slightly underestimated, with errors of approximately 1 dB at low frequencies and less than 2 dB at high frequencies.

The overall tendencies remained essentially unchanged for other microphone spacings. For shorter spacings, the angular error at 16 kHz was approximately 2$^\circ$ smaller for TF6, while the level error showed almost no change. For the largest spacing examined, $d=0.2$ m, the angular error increased slightly, but the degradation was typically within about 1$^\circ$. Again, no noticeable change was observed in the level estimation error. 
These results suggest that the robustness provided by the tight-frame averaging process is largely preserved over a practical range of microphone spacings.

\section{Conclusions}

This paper investigated sound-intensity estimation using cardioid microphone pairs arranged in tight-frame configurations. A spherical-harmonic error analysis was developed to characterize the propagation of direction-dependent microphone errors through the reconstruction process. The geometry-dependent coefficient $G_n$ was introduced and combined with the measured directivity-error coefficients to define the effective leakage indicator $\Lambda(\omega)$.

Within the evaluated tight-frame configurations, greater directional redundancy generally reduced directivity-related leakage, with TF24 exhibiting the smallest leakage over a wide range of spherical-harmonic degrees. Simulations using measured DPA2012 directivity data showed that TF12 and TF24 maintained direction-estimation errors of only a few degrees up to approximately 16~kHz under the examined noise and microphone-spacing conditions.

These results clarify the error-suppression mechanism underlying the previously observed robustness of tight-frame-based diffuseness measurements. Experimental validation using prototype cardioid microphone pairs will be reported separately.

\section*{ACKNOWLEDGMENTS}
The author would like to thank Mr. Ohno, whose early work demonstrated the remarkable robustness of the C-C method at high frequencies, and Dr. Kashiwazaki, who  recognized the relevance of tight-frame theory to directional sound-field measurements.
This research was supported by JSPS KAKENHI Grant No. JP24K03222.


\end{document}